\documentclass[sigconf]{acmart}

\acmConference[SPLC'19]{23rd International Systems and Software Product Line Conference}{9--13 September, 2019}{Paris, France}

\usepackage[T1]{fontenc} % textpipe
\usepackage{amsmath}
\usepackage{amssymb}

\usepackage{multirow}
\usepackage{multicol}
\usepackage{minibox}
\usepackage{todonotes}
\usepackage{listings}
\usepackage{balance}
\usepackage{hyperref}
\usepackage{graphicx}
\setcopyright{none}

\usepackage{pdfpages}
\usepackage{subfig}

\usepackage{algorithm}
\usepackage{algorithmic}

\usepackage{tcolorbox}
\newcommand{\SumupBox}[1]{
\begin{tcolorbox}[colback=gray!30,colframe=black, rounded corners = all,boxsep=1pt]
	#1
\end{tcolorbox}
}

\usepackage{xspace}
\newcommand\etal[0]{\emph{et al.}\xspace}
\newcommand{\ie}{\emph{i.e.,}\xspace}
\newcommand{\eg}{\emph{e.g.,}\xspace}

\newcommand{\etc}{\emph{etc.}\xspace}
\newcommand{\wrt}{w.r.t.\xspace}

\newcommand{\td}[1]{}

\usepackage{stmaryrd}

\newif\ifdraft\drafttrue
\ifdraft
\newcommand\todos[1]{\todo[inline]{TODO (all): #1}}
\newcommand\ma[1]{\todo[color=green!40,inline]{TODO (Mathieu): #1}}

\newcommand\jmj[1]{\todo[color=yellow!40,inline]{TODO (JMJ): #1}}
\newcommand\rb[1]{\todo[color=blue!35,inline, caption={}]{TODO (Razieh): #1}}

\newcommand\lnb[1]{\todo[inline,color=red!20!blue!10,bordercolor=red!20!blue!40]{TODO (Leo): #1}}

\else
\newcommand\todos[1]{}
\newcommand\ma[1]{}

\newcommand\jmj[1]{}
\newcommand\rb[1]{}

\newcommand\lnb[1]{}
\fi
\begin{document}

%\includepdf[pages=-]{./Includes/cover_letter.pdf}
%\input{Includes/ToBeRemoved.tex}

%\title{Adversarial Constraints for Variability Models}
%\title{Towards Adversarial Configurations for Software Product Lines}
%\title{Generating Adversarial Configurations for Quality Assurance of Software Product Lines}
\title{Towards Quality Assurance of Software Product Lines with Adversarial Configurations}

\author{Paul Temple}
\affiliation{%
  \institution{NaDI, PReCISE, Faculty of Computer Science, University of Namur}
  \city{Namur}
  \country{Belgium}
}

\author{Mathieu Acher}
\affiliation{%
  \institution{Univ Rennes, IRISA, Inria, CNRS}
  \city{Rennes}
  \country{France}
}

\author{Gilles Perrouin}
\affiliation{
    \institution{NaDI, PReCISE, Faculty of Computer Science, University of Namur}
    \city{Namur}
    \country{Belgium}
}

\author{Battista Biggio}
\affiliation{%
  \institution{University of Cagliari}
  \city{Cagliari}
  \country{Italy}
}

\author{Jean-Marc J\'{e}z\'{e}quel}
\affiliation{%
  \institution{Univ Rennes, IRISA, Inria, CNRS}
  \city{Rennes}
  \country{France}
}

\author{Fabio Roli}
\affiliation{%
  \institution{University of Cagliari}
  \city{Cagliari}
  \country{Italy}
}

\renewcommand{\shortauthors}{Paul Temple \etal}

\begin{abstract}

Software product line (SPL) engineers put a lot of effort to ensure that, through the setting of a large number of possible configuration options, products are acceptable and well-tailored to customers' needs. Unfortunately, options and their mutual interactions create a huge configuration space which is intractable to exhaustively explore. Instead of testing all products, machine learning techniques are increasingly employed to approximate the set of acceptable products out of a small training sample of configurations. Machine learning (ML) techniques can refine a software product line through learned constraints and \textit{a priori} prevent non-acceptable products to be derived. In this paper, we use adversarial ML techniques to generate \textit{adversarial configurations} fooling ML classifiers and pinpoint incorrect classifications of products (videos) derived from an industrial video generator. Our attacks yield (up to) a 100\% misclassification rate and a drop in accuracy of 5\%. We discuss the implications these results have on SPL quality assurance.

\end{abstract}

%page number
\settopmatter{printfolios=true}

\maketitle

\section{Introduction}
\label{sec:intro}

\begin{quote}
    Testers don't like to break things; they like to dispel the illusion that things work.
    ~\cite{Kaner:2001:LLS:559964}
\end{quote}

Software Product Line Engineering (SPLE) aims at delivering \textit{massively customized} products within shortened development cycles \cite{pohl-etal2005,clements2001}.
To achieve this goal, SPLE systematically reuses software assets realizing the functionality of one or more \textit{features}, which we loosely define as units of variability.
Users can specify products matching their needs by selecting/deselecting the features and provide additional values for their attributes. 
Based on such \textit{configurations}, the corresponding products can be obtained as a result of the product derivation phase. 
A long-standing issue for developers and product managers is to gain confidence that all possible products are functionally viable, \eg all products compile and run. This is a hard problem, since modern software product lines (SPLs) can involve thousands of features and practitioners cannot test all possible configurations and corresponding products due to combinatorial explosion. Research efforts rely on variability models (\eg feature diagrams) and solvers (SAT, CSP, SMT) to compactly define how features can and cannot be combined~\cite{batory2005,schobbens2007, berger2013,benavides2010}. Together with advances in model-checking, software testing and program analysis techniques, it is conceivable to assess the functional validity of configurations and their associated combination of assets within a product of the SPL~\cite{DBLP:conf/pldi/BoddenTRBBM13,DBLP:journals/fac/StruberRACTP18,classen2010,classen2011,DBLP:journals/jlp/BeekFGM16,DBLP:conf/icse/NadiBKC14}.\\
Yet, when dealing with qualities on the derived products (performance, costs, \etc) several unanswered challenges remain from the specification of feature-aware quantities to the best trade-offs between products and family-based analyses (\eg \cite{Legay:2017:QRP:3023956.3023970,terBeek:2019:QVM:3302333.3302349}). In our industrial case-study, the MOTIV video generator~\cite{galindoISSTA2014}, one can approximately generate $10^{314}$ video variants. Furthermore, it takes about $30$ minutes to create a new video: a non-acceptable (\eg a too noisy or dark) video can lead to a tremendous waste of resources. A promising approach is to sample a number of configurations and predict the quantitative or qualitative properties of the remaining configurations using Machine Learning (ML) techniques~\cite{SGKA:ESECFSE15,guo2015,guo2013,DBLP:conf/isola/BeekFGS16,siegmund2013,DBLP:conf/sigsoft/OhBMS17,temple:hal-01323446}.
These techniques create a predictive model (a classifier) from such sampled configurations and infer the properties of yet unseen configurations with respect to their distribution's similarity. This way, unseen configurations that do not match specific properties can be automatically discarded and constraints can be added to the feature diagram in order to avoid them permanently~\cite{DBLP:journals/software/TempleAJB17,temple:hal-01323446}.
However, we need to trust the ML classifier~\cite{barreno2006can,nelson2008} to avoid costly misclassifications. In the ML community, it has been demonstrated that some forged instances, called \emph{adversarial}, can fool a given classifier \cite{biggio2018wild}. \textit{Adversarial machine learning} (advML) thus refers to techniques designed to fool (\eg \cite{biggio2013poisoning,biggio2013evasion,nelson2008}), evaluate the security (\eg \cite{biggio2014security}) and even improve the quality of learned classifiers \cite{gan2014}. Our overall goal is to study how advML techniques can be used to assess quality assurance of ML classifiers employed in SPL activities.
In this paper, we design a generator of adversarial configurations for SPLs and measure how the prediction ability of the classifier is affected by such \textit{adversarial} configurations.
We also provide scenarios of usage of advML for quality assurance of SPLs. We discuss how adversarial configurations raise questions about the quality of the variability model or the testing oracle of SPL's products.
This paper makes the following contributions:
\begin{enumerate}
    \item An adversarial attack generator, based on evasions attacks and dedicated to SPLs;
    \item An assessment of its effectiveness and a comparison against a random strategy, showing that up to $100\%$ of the attacks are valid with respect to the variability model and successful in fooling the prediction of acceptable/non-acceptable videos, leading to a 5\% loss in accuracy;
    \item A qualitative discussion on the generated adversarial configurations \wrt to the classifier training set, its potential improvement and the practical impact of advML in the quality assurance workflow of SPLs.
    \item The public availability of our implementation and empirical results at \url{https://github.com/templep/SPLC_2019}
\end{enumerate}
The rest of this paper is organized as follows:
Section~\ref{sec:background} presents the case study and gives background information about ML and advML;
Section~\ref{sec:evasion_attack} shows how advML is used in the context of MOTIV;
Section~\ref{sec:eval} describes experimental procedures and discusses results;
Section~\ref{sec:threats} and~\ref{sec:discuss} presents some potential threats that could mitigate our conclusions and propose qualitative discussions about how adversarial configurations could be leveraged for SPLs developers.
Section~\ref{sec:related_work} covers related work and Section \ref{sec:conclusion} wraps up the paper with conclusions.

\section{Background}
\label{sec:background}

\subsection{Motivating case: MOTIV generator}
\label{sec:video_gen}

%We consider a representative SPL of our problem, an industrial video generator developed in the context of an industrial project called MOTIV\footnote{for simplicity, we will call the video generator MOTIV in the remaining of the paper.} (more details can be found in~\cite{galindoISSTA2014,temple:hal-01323446,mauricio2018}).
%The goal of MOTIV is to produce synthetic videos out of a high-level, textual specification.
%Generated videos are used to benchmark computer vision (CV) based systems under various conditions.
%For instance, one video may simulate fog, noise created by camera sensors, different illumination conditions (\eg day or night), \etc
%\gpe{day/night? adverse weather? etc. Give also examples of subjects, cars, people etc. to illustrate the context...}
%Variability management is crucial to produce a diverse yet realistic set of video variants, in a controlled and automated way. 

MOTIV is an industrial video generator which purpose is to provide synthetic videos that can be used to benchmark computer vision based systems.
Video sequences are generated out of configurations specifying the content of the scenes to render~\cite{temple:hal-01323446}.
MOTIV relies on a variability model that documents possible values of more than $100$ configuration options, each of them affecting the \textit{perception} of generated videos and the achievement of subsequent tasks, such as recognizing moving objects. Perception's variability relates to changes in the background (\eg being a forest or buildings), objects passing in front of the camera (with varying distances to the camera and different trajectories), blur, \etc There are $20$ Boolean options, $46$ categorical (encoded as enumerations) options (\eg to define predefined trajectories) and $42$ real-value options (\eg dealing with  blur or noise).
Precisely, enumerations contain about $7$ elements each and, in average, real-value options vary between $0$ and $27.64$ with a precision of $10^{-5}$.
Excluding (very few) constraints in the variability model, we over-estimate the video variants' space size: $2^{20}*7^{46}*((0-27.64)*10^{5})^{42} \approx 10^{314}$. Concretely, MOTIV takes as input a text file describing the scene to be captured by a synthetic camera as well as recording conditions.
Then, Lua~\cite{Ierusalimschy:2006:PLS:1200583} scripts are called to compose the scene and apply desired visual effects resulting in a video sequence.
To realize variability, the Lua code use parameters in functions to activate or deactivate options and to take into account values (enumerations or real values) defined into the configuration file.
A highly challenging problem is to identify feature values and interactions that make the identification of moving objects extremely difficult if not impossible. Typically, some of the generated videos contain too much noise or blur. In other words, they are \emph{not acceptable} as they cannot be used to benchmark object tracking techniques.
Another class of non-acceptable videos is composed of the ones in which pixels value do not change, resulting in a succession of images for which all pixels have the same color: nothing can be perceived.  As mentioned in Section~\ref{sec:intro}, non-acceptable videos represent a waste of time and resources: $30$ minutes per video, not including to run benchmarks related to object tracking (several minutes depending on the computer vision algorithm). We therefore need to constraint our variability model to avoid such cases.

\subsection{Previous work: ML and MOTIV}

\begin{figure}
    \centering
    \includegraphics[scale=0.32]{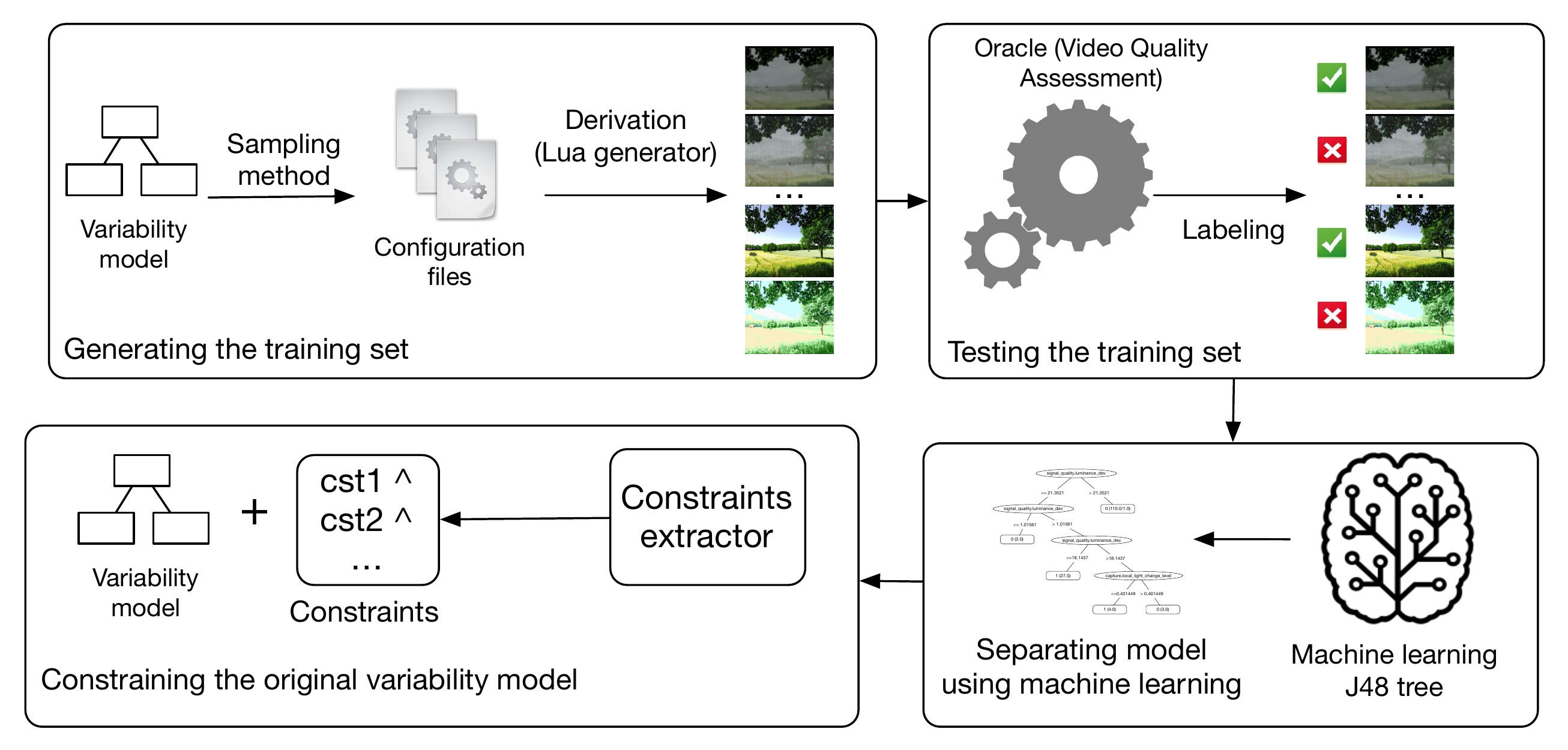}
    \vspace*{-4mm}
    \caption{Refining the variability model of MOTIV video generator via an ML classifier.}
    \vspace*{-4mm}
    \label{fig:ML_framework}
\end{figure}

We previously used ML classification techniques to predict the acceptability of unseen video variants~\cite{temple:hal-01323446}. We summarise this process in Figure~\ref{fig:ML_framework}.  

We first sample valid configurations using a random strategy (see Temple~\etal~\cite{temple:hal-01323446} for details) and generate the associated video sequences. A computer program playing the role of \textit{a testing oracle} then labels videos as acceptable (in green) or non-acceptable (in red). This oracle implements image quality assessment~\cite{IQA} defined by the authors via an analysis of frequency distribution given by Fourier transformations. An ML classifier (in our case, a decision tree) can be trained on such labelled videos. ``Paths" (traversals from the top to the leaves) leading to non-acceptable videos can easily be transformed into new constraints and injected in the variability model. An ML classifier can make errors, preventing acceptable videos (false negatives) or allowing non-acceptable videos (false positives).
Most of these errors can be attributed to the confidence of the classifier coming from both its design (\ie the set of approximations used to build its decision model) and the training set (and more specifically the distribution of the classes). Areas of low confidence exist if configurations are very dissimilar to those already seen or at the frontier between two classes. We use advML to quantify these errors and their impact on MOTIV. 

\subsection{ML and advML}

\begin{figure}
\centering
\includegraphics[scale=0.35]{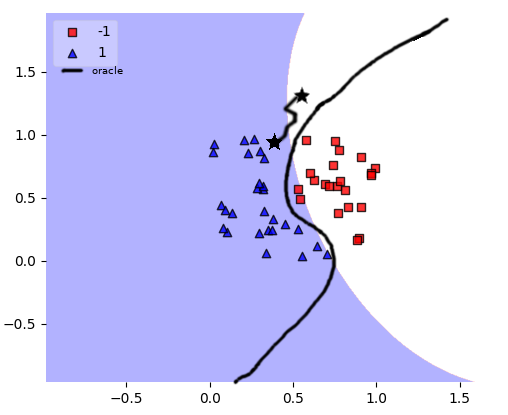}
\vspace*{-4mm}
\caption{Adversarial configurations (stars) are at the limit of the separating function learned by the ML classifier}
\label{fig:classif_errors}
\vspace*{-4mm}
\end{figure}

\emph{ML classification.}\label{subsec:ML} Formally, a classification algorithm builds a function $f: X \mapsto Y$ that associates a label in the set of predefined classes $y \in Y$ with configurations represented in a feature space (noted $x \in X$). In MOTIV, only two classes are defined: $Y = \{ -1, +1 \}$, respectively representing acceptable and non-acceptable videos.
$X$ represents a set of configurations and the configuration space is defined by configuration options of the underlying feature model (and their definition domain).
The classifier $f$ is trained on a data set $D$ constituted of a set of pairs ($x^t_i$, $y^t_i$) where $x_i \in X$ are a set of valid configurations from the variability model and $y_i \in Y$ their associated labels. To label configurations in $D$, we use an oracle (see Figure~\ref{fig:ML_framework}). Once the classifier is trained, $f$ induces a separation in the feature space (shown as the transition from the blue/left to the white/right area in Figure~\ref{fig:classif_errors}) that mimics the oracle: when an unseen configuration occurs, the classifier determines instantly in which class this configuration belongs to. Unfortunately, the separation can make prediction errors since the classifier is based on statistical assumptions and a (small) training sample. We can see in Figure~\ref{fig:classif_errors} that the separation diverges from the solid black line representing the target oracle.  As a result, two squares are misclassified as being triangles.  Classification algorithms realise trade-offs between the necessity to classify the labelled data correctly, taking into account the fact that it can be noisy or biased and its ability to generalise to unseen data. Such trade-offs lead to approximations that can be leveraged by adversarial configurations (shown as stars in Figure~\ref{fig:classif_errors}). 

\emph{AdvML and evasion attacks.} According to Biggio~\etal~\cite{biggio2018wild}, deliberately attacking an ML classifier with crafted malicious inputs was proposed in 2004.
%Since then, it emerged as a sub-discipline of machine learning called adversarial machine learning. 
Today, it is called adversarial machine learning and can be seen as a sub-discipline of machine learning.
Depending on the attackers' access to various aspects of the ML system (dataset, ability to update the training set) and their goals, various kinds of attacks~\cite{biggio2013poisoning,biggio2012poisoning, biggio2013evasion,biggio2014security,biggio2014pattern} are available: they are organised in a taxonomy~\cite{barreno2006can,biggio2018wild}. In this paper, we focus on \textit{evasion attacks}: these attacks move labelled data to the other side of the separation (putting it in the opposite class) via successive modifications of features' values. Since areas close to the separation are of low confidence, such adversarial configurations can have a significant impact if added to the training set. To determine the direction to move the data towards the separation, a gradient-based method has been proposed by Biggio~\etal~\cite{biggio2013evasion}. This method requires the attacked ML algorithm to be differentiable. One of such differentiable classifiers is the Support Vector Machine (SVM), parameterizable with a kernel function\footnote{most common functions are linear, radial based functions and polynomial}. %We present in the next section how we adapted this method to the MOTIV case.

\section{Evasion attacks for MOTIV}
\label{sec:evasion_attack}

\subsection{A dedicated Evasion Algorithm}

\begin{algorithm}
\caption{Our algorithm conducting the gradient-descent evasion attack inspired by~\cite{biggio2013evasion}}
\label{alg:attack}
    \textbf{Input: } \textit{$x^0$}, the initial configuration; \textit{t}, the step size; \textit{nb\_disp}, the number of displacements; \textit{g}, the discriminant function\\
    \textbf{Output: }\textit{$x^{*}$}, the final attack point
    \begin{algorithmic}
        \STATE(1) m = 0;
        \STATE(2) Set $x^0$ to a copy of a configuration of the class from which the attack starts;
        \WHILE{$m < \textit{nb\_disp}$}
            \STATE(3) m = m+1;
            \STATE(4) Let $\nabla F(x^{m-1})$ a unit vector, normalisation of $\nabla g(x^{m-1})$;
            \STATE(5) $x^m$ = $x^{m-1} - t\nabla F(x^{m-1})$;
            
        \ENDWHILE
        \STATE(6) return $x^{*}$ = $x^m$;
    \end{algorithmic}
\end{algorithm}

Algorithm~\ref{alg:attack} presents our adaptation of Biggio~\etal's evasion attack~\cite{biggio2013evasion}. First, we select an initial configuration to be moved ($x^0$): selection tradeoffs are discussed in the next section. Then, we need to set the step size ($t$), a parameter controlling the convergence of the algorithm. Large steps induce difficulties to converge, while small steps may trap the algorithm in a local optimum.  While the original algorithm introduced a termination criterion based on the impact of the attack on the classifier between each move (if this impact was smaller than a threshold $\epsilon$, the algorithm stopped; assuming an optimal attack) we fixed the maximal number of displacements $nb\_disp$ in advance. This allows for a controllable computation budget, as we observed that for small step sizes the number of displacements required to meet the termination criterion was too large. The function $g$ is the discriminant function and is defined by the ML algorithm that is used. It is defined as $g: X \mapsto \mathbb{R}$ that maps a configuration to a real number.
In fact, only the sign of $g$ is used to assign a label to a configuration $x$.
Thus, $f: X \mapsto Y$ can be decomposed in two successive functions: first $g: X \mapsto \mathbb{R}$ that maps a configuration to a real value and then $h: \mathbb{R} \mapsto Y$ with $h= sign(g)$.
However, $|g(x)|$ (the absolute value of $g$) intuitively reflects the confidence the classifier has in its assignment of $x$. $|g(x)|$ increases when $x$ is far from the separation and surrounded by other configurations from the same class and is smaller when $x$ is close to the separation.
%Intuitively, this number reflects the confidence that the classifier has in the fact that a configuration belongs to a specific class since $|g(x)|$ (the absolute value of $g$) increases when $x$ is far from the separation and surrounded by other configurations from the same class. Knowing the sign of $g$ is sufficient to associate a label to a configuration $x$ in our case (since $Y = \{+1 ; -1\}$).
Using this discriminant function has been proposed by Biggio~\etal~\cite{biggio2013evasion} and should not be confused with the unrelated discriminator component of GANs by Goodfellow~\etal~\cite{gan2014}. In GANs, the discriminator is part of the ``robustification process". It is an ML classifier striving to determine whether an input has been artificially produced by the other GANs' component, called the generator. Its responses are then exploited by the generator to produce increasingly realistic inputs. %that cannot be distinguished anymore. 
In this work, we only generate adversarial configurations, though GANs are envisioned as follow-up work.

Concretely, the core of the algorithm consists of the $while$ loop that iterates over the number of displacements. Statement (4) determines the direction towards the area of maximum impact with respect to the classifier (explaining why only a unit vector is needed). $\nabla g(x^{m-1})$ is the gradient of $g(x^{m-1})$ and the direction of interest towards which the adversarial configuration should move. This vector is then multiplied by the step size $t$ and subtracted to the previous move (5). The final position is returned after the number of displacements has been reached. For statements (4) and (5) we simplified the initial algorithm~\cite{biggio2013evasion}: we do not try to mimic as much as possible existing configurations as we look forward to some diversity. In an open ended feature space, gradient can grow indefinitely possibly preventing the algorithm to terminate. Biggio~\etal~\cite{biggio2013evasion} set a maximal distance representing a boundary of the feasible region to keep the exploration under control. In MOTIV, this boundary is represented by the hard constraints in the variability model. Because of the heterogeneity of MOTIV features, cross-tree constraints and domain values are difficult to specify and enforce in the attack algorithm. SAT/SMT solvers would slow down the attack process. We only take care of the type of feature values (natural integers, floats, Boolean). For example, we reset to zero natural integer values that could be negative due to displacements or we ensure that Boolean values are either 0 or 1.

%only Algorithm~\ref{alg:attack} is a generator of adversarial configurations, the discrimination function is inherent to a classifier (\ie its decision process) and is used to conduct the attack proposed by Biggio~\etal~\cite{biggio2013evasion}.

% however, they are unrelated and used in different context.
%As we just described, here the discriminant function is the way a classifier can predict whether an example belongs to one class or an other.

%guides the exploration towards the separation taking into account the density of labelled configurations to focus on areas of low confidence (where the density is low as well).
%Thus, $g: X \mapsto [-1,+1]$ is a combination of the distance to the separation and of the confidence of the classifier in its prediction. 
%For instance, if a configuration $x$ is far from the separation and surrounded by other configurations from class +1, $g(x)$ will be close to $+1$ as it is very unlikely that a configuration from class $-1$ while lies in this area. Going from $g$ to $Y$ (\ie the label associated to a configuration) for a specific configuration is simply done by considering the sign of $g$. 

%The algorithm works as follows.  After initialising the number of displacement counter $m$ and the initial configuration, we enter in a loop until $nb_{disp}$ has been reached. 

\begin{figure*}
    \centering
   \begin{minipage}{.3\linewidth}
       \subfloat[]{
           \includegraphics[scale=0.10]{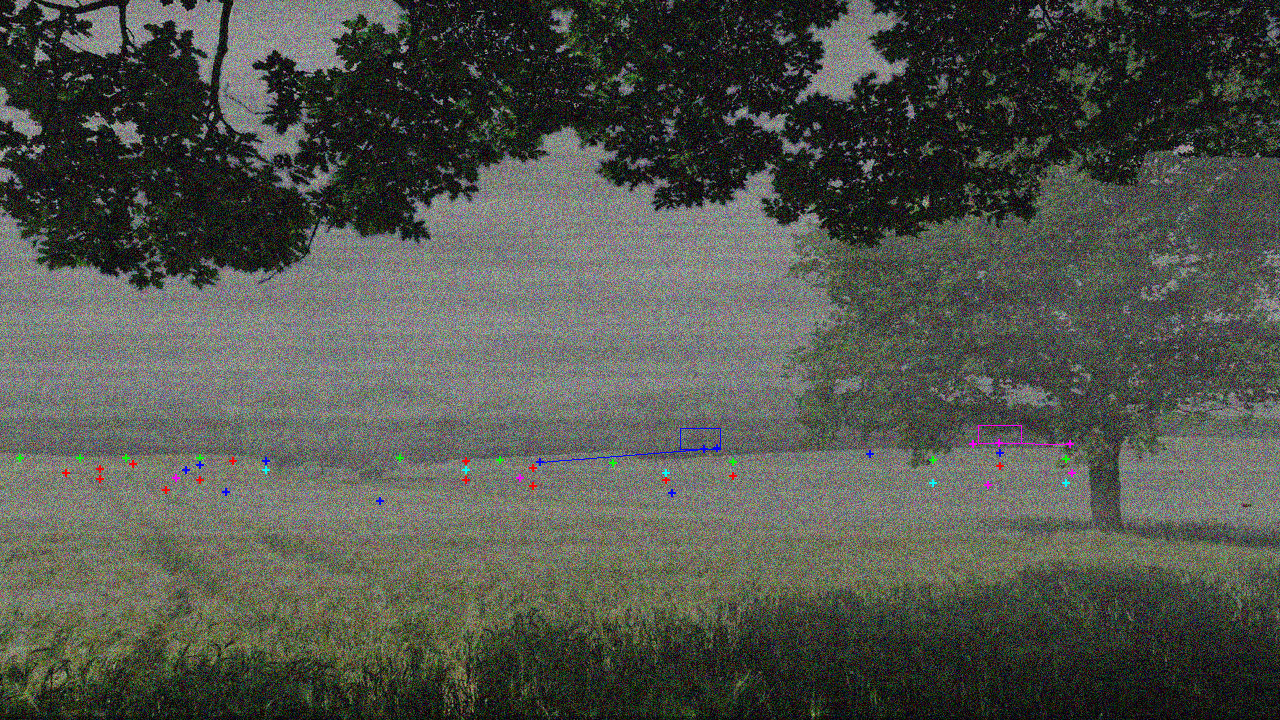}
           \label{img:ex1}
       }
   \end{minipage}
   \begin{minipage}{.3\linewidth}
       \subfloat[]{
           \includegraphics[scale=0.10]{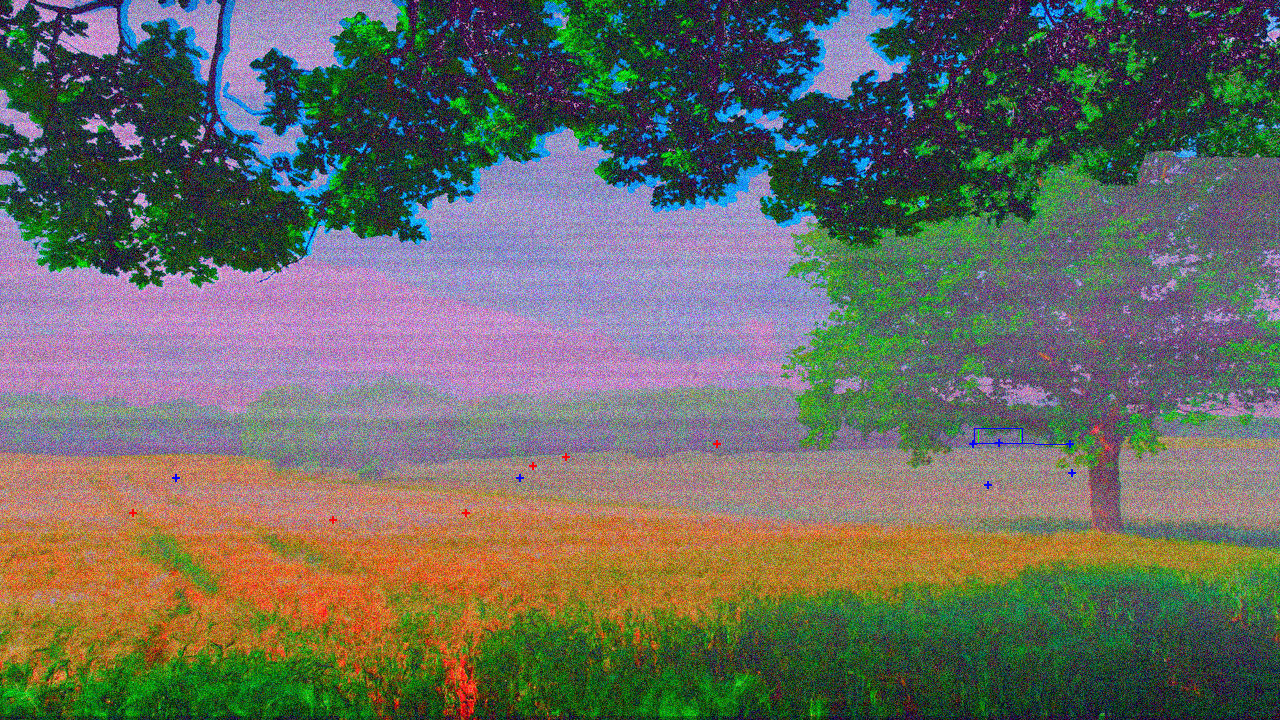}
           \label{img:ex2}
       }
   \end{minipage}
   \begin{minipage}{.3\linewidth}
       \subfloat[]{
           \includegraphics[scale=0.10]{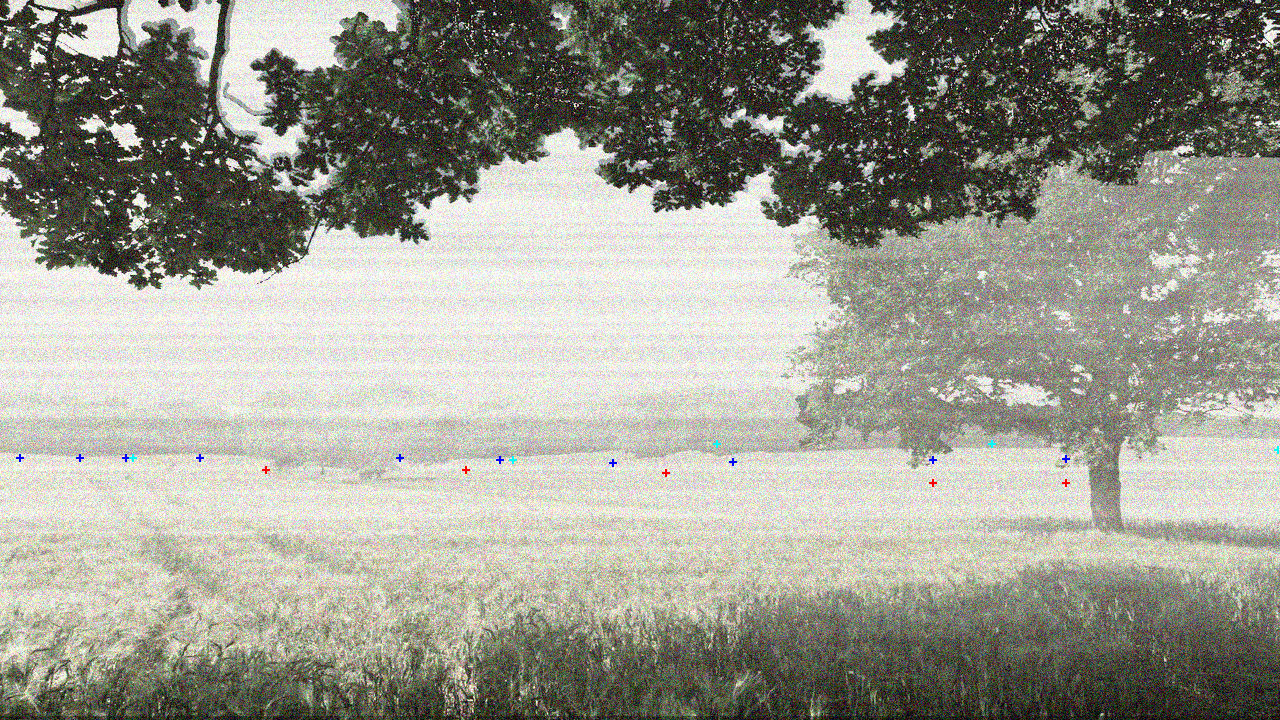}
           \label{img:ex3}
       }
   \end{minipage}
   \begin{minipage}{.3\linewidth}
       \subfloat[]{
           \includegraphics[scale=0.10]{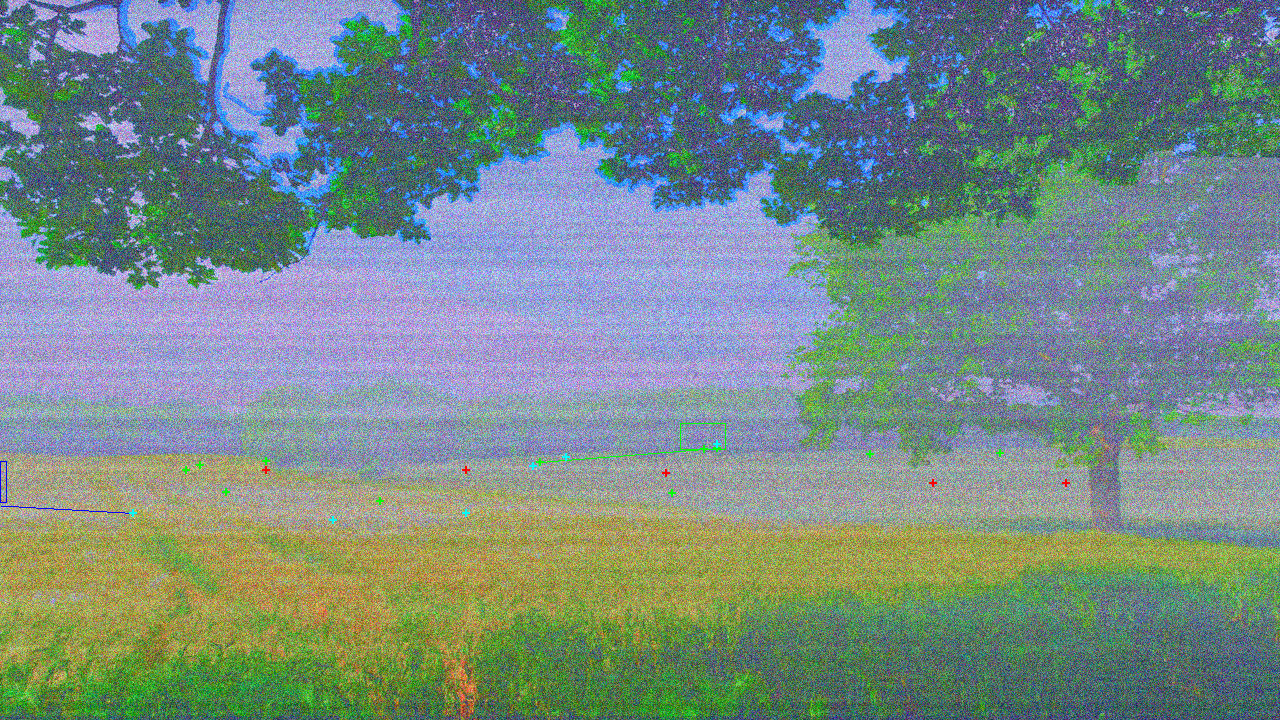}
           \label{img:ex4}
       }
   \end{minipage}
   \begin{minipage}{.3\linewidth}
       \subfloat[]{
           \includegraphics[scale=0.10]{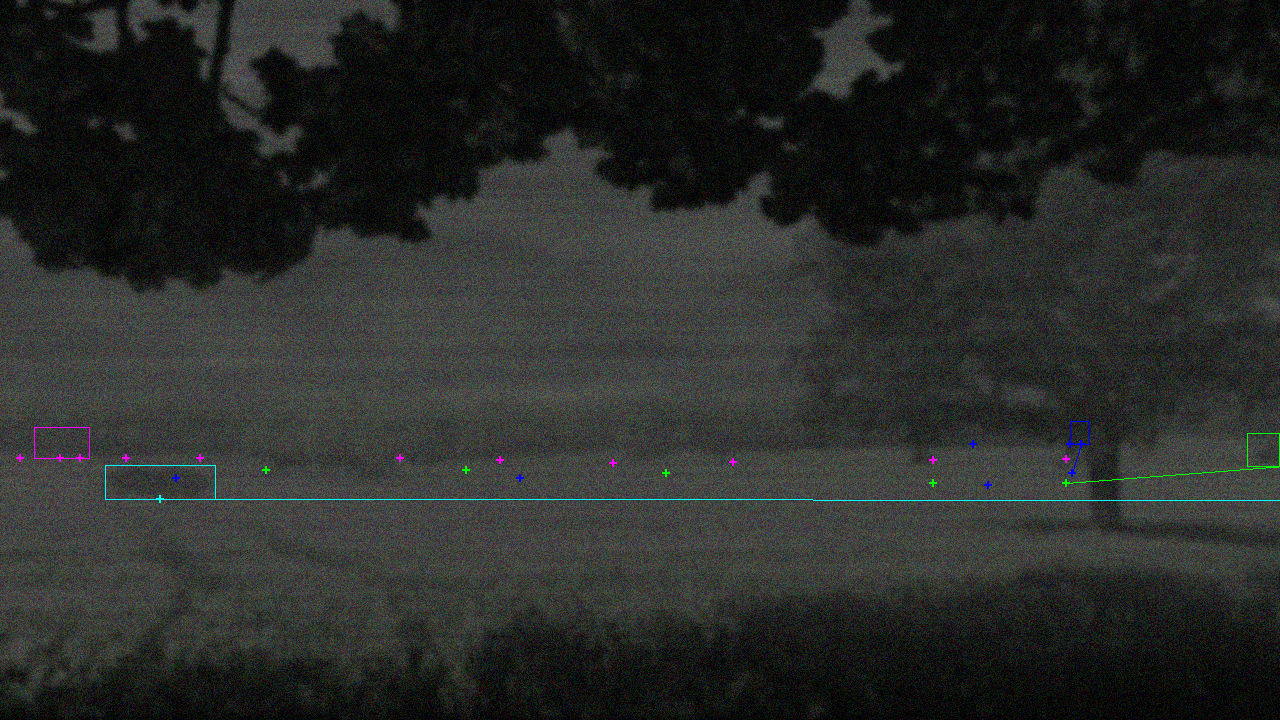}
           \label{img:ex6}
       }
   \end{minipage}
   \vspace*{-4mm}
   \caption{Examples of generated videos using evasion attack}
   \label{img:examples_gen_vid}
   \vspace*{-4mm}
\end{figure*}

%Thus, we only took care of the structural constraints (such as mutual exclusions and feature value intervals) in our implementation:  when a displacement would exceed the specified interval for a feature value, we set it to closest acceptable bound. and discuss this choice in Section~\ref{sec:discuss}. 

As introduced in Section~\ref{sec:background}, decision trees are not directly compatible with evasion attacks as the underlying mathematical model is highly non-linear making it non-derivable (forbidding to compute a gradient).  We learn another classifier (\ie a Support Vector Machine) on which we can perform evasion attacks directly~\cite{biggio2013evasion,biggio2018wild}.
We rely upon evidence that attacks conducted on a specific ML model can be transferred to others~\cite{Demontis2019Transfer,Demontis2018Transfer,brownadversarial,brown2017adversarial}.%, which we rely upon, that attacks can be transferred from a specific ML model towards others.
%There is evidence that attacks conducted on a specific ML model can be transferred towards other models~\cite{Demontis2019Transfer,Demontis2018Transfer,brownadversarial,brown2017adversarial}, which we rely upon.  

%Demontis~\etal~\cite{Demontis2019Transfer,Demontis2018Transfer} studied the fact that attacks conducted on a specific ML model can be transfered towards other models. In the same vein, %Brown~\etal~\cite{brownadversarial,brown2017adversarial} showed that universal attacks can be conducted towards many different kind of ML algorithms such that they can all be fooled by the same input.

\subsection{Implementation}

We implemented the above procedure in Python 3 (scripts available on the companion website). Figure~\ref{img:examples_gen_vid} depicts some images of videos generated out of adversarial configurations.
%MOTIV's variability model describes heterogeneous features: some are Boolean, other are real or even enumerations.

MOTIV's variability model embeds enumerations which are usually encoded via integers.
The main difference between the two is the logical order that is inherent to integers but not encoded into enumerations. As a result, some ML techniques have difficulties to deal with them. 
The solution is to ``dummify" enumerations into a set of Boolean features, which truth values take into account exclusion constraints in the original enumerations. Conveniently, Python provides the \emph{get\_dummies} function from the pandas library which takes as input a set of configurations and feature indexes to dummify.
%and returns a set of Boolean features whose number of element is the maximum litteral index encountered for each feature index. 
For each feature index, the function creates and returns a set of Boolean features representing the literals indexes encountered while running through given configurations: if the \textit{get\_dummies} function detects values in the integer range $[0,9]$ for a feature associated to an enumeration, it will return a set of $10$ Boolean features representing literals indexes in that range. It also takes care or preserving the semantics of enumerations. However, dummification is not without consequences for the ML classifier. First, it increases the number of dimensions: our $46$ initial enumerations would be transformed into $145$ features. Doing so may expose the ML algorithm to the \textit{curse of dimensionality}~\cite{Bellman:1957}: as the number of features increases in the feature space, configurations that look alike (\ie with close feature values and the same label) tend to go away from each other, making the learning process more complex. This curse has also been recognised to have an impact on SPL activities \cite{davril:hal-01243571}. Dummification implies that we will operate our attacks in a feature space that is the \textit{essentially different} than the one induced by the real SPL. This means that we need to transpose the generated attacks in the dummified feature space back to the original SPL one, raising one main issue: there is no guarantee that an attack relevant in the dummified space is still efficient in the reduced original space (the separation may simply not be the same). 
%Additionally, gradient methods operate per feature only, meaning that it ignores exclusion constraints in dummified enumerations, yielding invalid configurations that would need to be ``fixed" when transposing the configurations in the original space, potentially further deriving these adversarial configurations from the optimum computed by the gradient method. 
Additionally, gradient methods operate per feature only, meaning that exclusion constraints in dummified enumerations are ignored.
That is, when transposed back to the original configuration space, invalid configurations would need to be ``fixed", potentially putting these adversarial configurations away from the optimum computed by the gradient method. 
For all these reasons, we decided to operate on the initial feature space, acknowledging the threat of considering enumerations as ordered. We conducted a preliminary analysis\footnote{available on the companion webpage: \url{https://github.com/templep/SPLC_2019}} that showed that the order of the importance of the features were kept whether we use a dummified or the initial feature space. So this threat is minor in comparison of the pitfalls of dummification.
We do not make any further distinctions between the two terms since we use them without making any transformations.

Since the attack cannot be conducted directly on decision trees, we decided to learn another classifier first. We chose support vector machines with a linear kernel since it was faster according to a preliminary experiment. Scripts as well as data used to compare predictions can be found on the companion webpage.  %preliminary experiment showed that linear and radial based functions give similar results in their prediction and we choose a linear kernel as it is faster to compute than the other.

\section{Evaluation}
\label{sec:eval}

\subsection{Research questions}

We address the following research questions:
\newline
\hspace*{2mm}\textbf{RQ1}: \emph{How effective is our adversarial generator to synthesize adversarial configurations?}
Effectiveness is measured through the capability of our evasion attack algorithm to generate configurations that are misclassified:
\begin{itemize}
    \item \textbf{RQ1.1}: Can we generate adversarial configurations that are wrongly classified?
    \item \textbf{RQ1.2}: Are all generated adversarial configurations valid \wrt constraints in the VM?
    \item \textbf{RQ1.3}: Is using the evasion algorithm more effective than generating adversarial configurations with random modifications?
    \item \textbf{RQ1.4}: Are attacks effective regardless of the targeted class?
\end{itemize}

\textbf{RQ2}: \emph{What is the impact of adding adversarial configurations to the training set regarding the performance of the classifier?}
The intuition is that adding adversarial configurations to the training set could improve the performance of the classifier when evaluated on a test set.

% Taking a step backwards, we can wonder what does the existence of adversarial configurations suggest in regards to an SPL like the MOTIV generator. 

\subsection{Evaluation protocol}
\label{subsec:eval_proto}

Our evaluation dataset is composed of $4,500$ randomly sampled and valid video configurations, that we used in previous work~\cite{temple:hal-01323446}. We selected $500$ configurations to train the classifier keeping a similar representation of non-acceptable configurations ($10\%$, \ie $\approx 50$ configurations) compared to the whole set. The remaining $4,000$ configurations are used as a test set and also have a similar representation regarding acceptable/non-acceptable configurations. This setting contrast with a common practice of using a high percentage (\ie around 66\%) of available examples to train the classifier. However, due to the low number of non-acceptable configurations, such a setting is impossible. $k$-fold cross-validation is another common practice used when few data points are available for training ($4,500$ configurations is an arguably low number with respect to the size of the variant space) but it is used to validate/select a classifier when several are created (which is not our case here) and in addition, separating our $4,500$ configurations into smaller sets is likely to create a lot of sets without any non-acceptable configurations.
None of these practices seem to be adapted to our case.
%While proposed when few data points are available for training ($4,500$ configurations is an arguably low number with respect to the size of the variant space), k-fold cross-validation is not the solution here as the test set will be very small and thus is likely to not include any non-acceptable configurations.

The key point is that only about 10\% of configurations are non-acceptable. This is a ratio that we cannot control exactly as it depends from the targeted non-functional property. In order to reduce imbalance, several data augmentation techniques exist like SMOTE~\cite{SMOTE}. Usually, they create artificial configurations while maintaining the configurations' distribution in the feature space.
In our case, we compute the centroid between two configurations and use it as a new configuration.
% In particular, one technique computes the centroid between two configurations and uses it as a new configuration.
% The centroid method has pros and cons: centroids are easy and quick to compute, new configurations tend to follow the same distribution as they result in more densely populated clusters (and on rare occasions, make clusters expand a little bit). % However, new configurations may not be realistic, they do not provide so much diversity since centroids, by definition, lie in the middle of the cluster of points.
 % Since our goal is only to limit imbalance in the available configurations, this technique seems appropriate while maintaining the initial distribution of configurations. 
 Thanks to the centroid method, we can bring perfect balance between the two classes (\ie $50\%$ of acceptable configurations and non-acceptable configurations).
Technically, we compute how many configurations are needed to have perfectly balanced sets (\ie training and test sets): We select randomly two non-acceptable configurations and compute the centroid between them, check that it is a never-seen-before configuration and adds it to the available configurations. The process is repeated until the number of configurations required is reached. Once a centroid is added to the set of available configurations, it is available as a configuration to create the next centroid. 

In the remainder, we present the results with both original and balanced data sets in order to assess whether the impact of class representation imbalance on adversarial attacks. We configured our evasion attack generator with the following settings: \textit{i)} we set the number of attacks points to generate $4000$ configurations for RQ1 and $25$ configurations for RQ2 as explained hereafter; \textit{ii)} considered step size ($t$) values are $\{10^{-6}$; $10^{-4}$; $10^{-2}$; $1$; $10^{2}$; $10^{4}$; $10^{6}\}$; \textit{iii)} the number of iterations is fixed to $20$, $50$ or $100$. To mitigate randomness, we repeat ten times the experiments. All results discussed in this paper can also be found on our companion webpage\footnote{\url{https://github.com/templep/SPLC_2019}}.

\subsection{Results}
%\begin{itemize}
%    \item RQ1:  How does it compare to a random baseline?
%        \begin{itemize}
%            \item Number of successful attacks over $4,000$ generations
%            \item Number of valid adv. configs over $4,000$ generations
%        \end{itemize}
%    \item RQ2: What is the impact of adding adv. configs in the training set?
%        \begin{itemize}
%            \item Do they improve the classifier performances?
%            \item What are the impacts on the resulting decision tree? (structure, readibility of extracted constraints)
%        \end{itemize}
%\end{itemize}

\subsubsection{\textbf{RQ1: How effective is our adversarial generator to synthesize adversarial configurations?}}\hfill

%We assess  by the ML classifier following Algorithm~\ref{alg:attack}. % inspired from Biggio~\etal~\cite{biggio2013evasion}).
To answer this question, we assess the number of wrongly classified adversarial configurations over $4,000$ generations (and about $7,000$ configurations when the training set is balanced) and compare them to a random baseline: there is to the best of our knowledge no comparable evasion attack.
%As, to the best of our knowledge, this evasion attack method is new for SPL, our baseline is formed by configurations that would be moved randomly in the configuration space.

\textbf{RQ1.1: Can we generate adversarial configurations that are \allowbreak wrongly classified?}
% First, we would like to assess that our adversarial configurations are effective.

\begin{figure*}
    \subfloat[Number of misclassified adversarial configurations (20 displacements) \label{subfig:adv_attack_20}]{%
    \includegraphics[width=.485\linewidth]{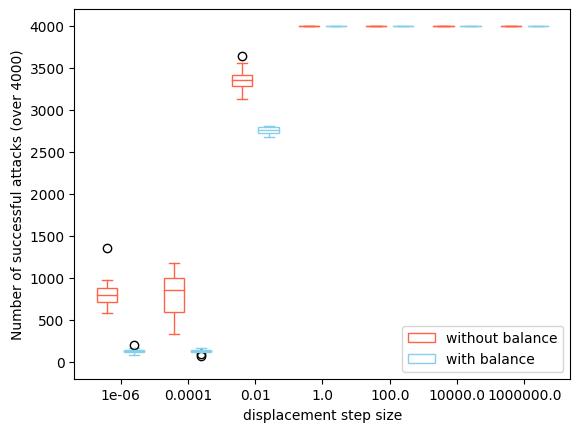}
    }
    \qquad
%    \subfloat[Number of successful adversarial attacks after 50 displacements \label{subfig:adv_attack_50}]{%
%    \includegraphics[scale=.4]{./Images/result_exec_attack/adv_attack/result_50_disp_steps}
%    }
    \subfloat[Number of misclassified adversarial configurations (100 displacements) \label{subfig:adv_attack_100}]{%
    \includegraphics[width=.47\linewidth]{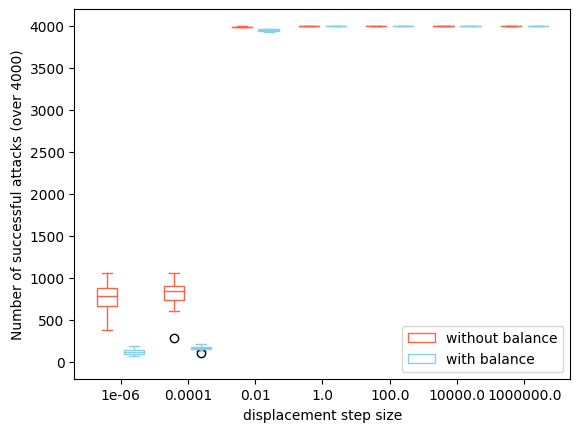}
    }
    \caption{Number of successful attacks on class \textit{acceptable}; X-axis represents different step size values $t$ while Y-axis is the number of misclassified adversarial configurations by the classifier. For each $t$ value, results with balanced and imbalanced training set are shown (respectively in blue and orange).}
    \label{fig:attack_success_acc}
\end{figure*}
%We first make vary $t$ (the number of displacements) as described in Section~\ref{subsec:eval_proto}. 
For each run, a newly created adversarial (\ie after $nb\_disp$ is reached) configuration is added to the set of initial configurations that can be selected to start an evasion attack. We therefore give a chance to previous adversarial configurations to continue further their displacement towards the global optimum of the gradient.

Figure~\ref{fig:attack_success_acc} shows box-plots resulting of ten runs for each attack setting. We also show results when the training set is imbalanced (\ie using the previous training set containing $500$ configurations with about $10\%$ of non-acceptable configurations) and when it is balanced (\ie increasing the number of non-acceptable configurations using the data augmentation technique described above).
% in order to flatten the impact of random choices made to create the initial position of adversarial attacks.
% \pt{I don't get the two following sentences}
% Note that, for each set of parameters, previous adversarial configurations were included in the set of potential candidates (\ie configurations in the initial training set with the label that should lead the attack, in this case "non-acceptable").
%Please note that, 
%to be taken again as a new adversarial configuration and continue the displacement process, going deeper into the attacked class (\ie acceptable).
 Both Figure~\ref{subfig:adv_attack_20} and Figure~\ref{subfig:adv_attack_100} indicate that we can always achieve $100\%$ of misclassified configurations with our attacks.
Regarding Figure~\ref{subfig:adv_attack_20}, all generated configurations become misclassified when step size is set to $1.0$ or a higher value. When $100$ displacements are allowed (see Figure~\ref{subfig:adv_attack_100}), the limit appears earlier, \ie when $t$ equals $0.01$.
Similar results can be obtained when the number of maximum displacements is set to $50$, the only difference is that with $t$ set to $0.01$ not all adversarial configurations are misclassified but about $3,900$ ($97.5\%$) when the training set is imbalanced and about $3,700$ ($92.5\%$) with a balanced set.
%Finally, Figure~\ref{subfig:adv_attack_100} reaches the maximum score at step size $0.001$.
%The fact that, at some point, we can find some combinations of step size and number of displacements that allow to generate only successful adversarial configurations is a bit surprising.
%In the end, evasion attacks can be parameterized such that all generated adversarial configurations are misclassified.
%combinations of step size and number of maximum displacements can be found such that all generated adversarial configurations are successful.
 
 \emph{Discussion.} Increasing the number of displacements require lower step sizes to reach the misclassification goal but it comes at the cost of more computations. However, increasing the number of displacements when the step size is already large results in incredibly large displacements, leading to invalid configurations in the MOTIV case. %In MOTIV, this is not tolerated because feature values need to belong to specific intervals to be part of a valid configuration (see below).  
%However, the fact that with a higher number of displacements a smaller step size is needed seems natural.
%In the end, it seems that increasing the number of iterations will always give a 100\% of successful attacks at the cost of more computations.
% However, increasing the number of displacements when the step size is already large would result in incredibly large displacements. Without any restrictions it would be okay but we remind that MOTIV defines some constraints and ranges of values.

\textbf{RQ1.2: Are all generated adversarial configurations valid \wrt constraints in the VM?}

As discussed in Section \ref{sec:evasion_attack}, we perform a basic type check on features. However, this check does not cover specific constraints such as cross-tree ones. To ensure the full compliance of our adversarial configurations, we run the analysis realised by the MOTIV video generator. This includes, amongst others, checking the correctness features values with respect to their specified intervals.   
%In the definition of MOTIV's VM, there were enumerations (converted into a set of Boolean features for the sake of adversarial attacks) which define a bounded set of possible choices.
%With displacements, it is possible that feature values go outside of the initial bounds.
%Because of the variety of constraints that had to be checked (\ie bounds, floats, non-negative values, cross-tree constraints, \etc), it is difficult to call a solver in the process that would respond quickly and on frequent occasions (since the adversarial attack is an iterative process).
%Still we were able to encode cross-tree constraints which were somehow already included in the configuration process.

%By design, our implementation checks boundary compliance of the $4,000$ generated adversarial configurations (see Section \ref{sec:evasion_attack}).
\begin{figure}
    \includegraphics[width=.95\linewidth]{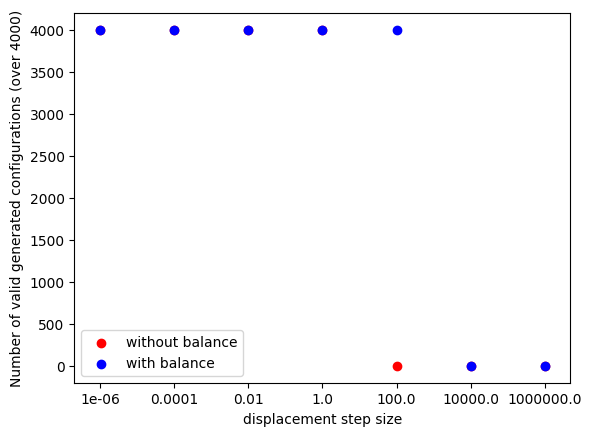}
    \caption{\label{fig:attack_valid_acc}Number of valid attacks on class \textit{acceptable}; X-axis represents different step size $t$ values; Y-axis reports the number of valid configurations. In red and blue are respective results with an imbalance and a balance training set in terms of classes representation.}
    % \label{subfig:valid_attack_20}
\end{figure}
%%%%% Old figure
%\begin{figure*}
%    \subfloat[20 displacements \label{subfig:valid_attack_20}]{%
%    \includegraphics[scale=.34]{./Images/result_valid_attack/adv_attack/result_valid_20_disp_steps}
%    }
%    \subfloat[50 displacements \label{subfig:valid_attack_50}]{%
%    \includegraphics[scale=.34]{./Images/result_valid_attack/adv_attack/result_valid_50_disp_steps}
%    }
%    \subfloat[100 displacements \label{subfig:valid_attack_100}]{%
%    \includegraphics[scale=.34]{./Images/result_valid_attack/adv_attack/result_valid_100_disp_steps}
%    }
%   \caption{Number of valid attacks on class -1; X-axis represents different step size $t$ values; Y-axis reports the number of valid configurations}
%    \label{fig:attack_valid_acc}
% \end{figure*}
Figure~\ref{fig:attack_valid_acc} shows on the X-axis the different step sizes while the Y-axis depicts the number of valid adversarial configurations \wrt constraints.
%It should be noted that the results are the same for $20$, $50$, or $100$ displacements -- we only depict the results for $20$ displacements. 
Regardless of the number of displacements and whether the training set is balanced, all results are the same except for Figure~\ref{fig:attack_valid_acc} that presents one difference for a displacement step size of $100$.
One possible explanation is that when the training set is balanced, more configurations can be taken as a starting point of the evasion algorithm: the gradient descent procedure might lead the current attack towards a slightly different area in which configurations remain valid.
Overall, regardless of the number of authorized displacements, we can see a clear drop of valid configurations from $4,000$ to $0$ between step size set to $1$ and $100$.

\emph{Discussion:} We thus can scope the parameters such that adversarial configurations are \emph{both successful and valid}: when step size is set to a value between $0.01$ and $1.0$, regardless of the number of displacements. Increasing the step size leads to non-valid configurations while with smaller step sizes, adversarial configurations have not moved enough to cross the separation of the classifier (leading to unsuccessful attacks).

\textbf{RQ1.3: Is using the evasion algorithm more effective than generating adversarial configurations with random displacements?}
Previous results of RQ1.1 and RQ1.2 show we are able to craft valid adversarial configurations that can be misclassified by the ML classifier, but is our algorithm better than a random baseline?
%To the best of our knowledge, we do not know any methods that try to challenge the ML classifier and exhibit interesting configurations that can be exploited to understand why and what makes the classifier fail.
%Because of that, we compare our adversarial approach to configurations that are randomly modified.
The baseline algorithm consists in: i) for each feature, choosing randomly whether to modify it; ii) choosing randomly to follow the slope of the gradient or going against it (the role of `-' of line 5 in Algorithm~\ref{alg:attack} that can be changed into a `+');
%a direction, in the feature space in which the modifications occur (it takes the role as the '-' of line 4 in Algorithm~\ref{alg:attack} that can be changed into a '+' corresponding to either following the slope of the gradient or going against it);
iii) choosing randomly a degree of displacement (corresponding to the slope of the gradient ($\nabla F(x^{m-1})$) of line 5 in Algorithm~\ref{alg:attack}).
Both the step size and the number of displacements are the same as in the previous experiments.

\begin{figure*}
    \subfloat[Number of successful random attacks after 20 displacements \label{subfig:rand_attack_20}]{%
    \includegraphics[width=.48\linewidth]{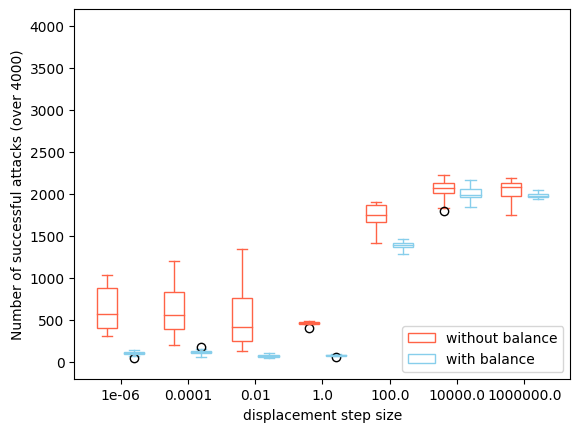}
    }
    \qquad
%    \subfloat[Number of successful random attacks after 50 displacements \label{subfig:rand_attack_50}]{%
%    \includegraphics[width=.45\linewidth,height=5cm]{./Images/result_exec_attack/random_attack/result_random_50_disp_steps_big}
%    }
    \subfloat[Number of successful random attacks after 100 displacements \label{subfig:rand_attack_100}]{%
   \includegraphics[width=.47\linewidth]{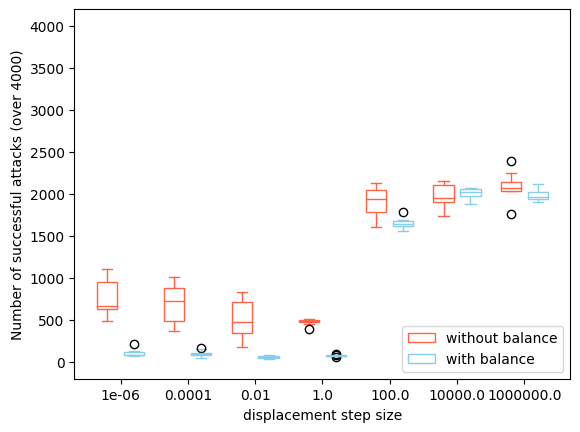}
    }
    \caption{Number of successful random attacks on class \textit{acceptable}; X-axis represents different step size values $t$ while Y-axis is the number of misclassified adversarial configurations by the classifier. In red and blue are respective results with an imbalance and a balance training set in terms of classes representation.}
    \label{fig:rand_attack_success_acc}
\end{figure*}
Figure~\ref{fig:rand_attack_success_acc} shows the ability of random attacks to successfully mislead the classifier.
%While adversarial attacks were able to reach a $100\%$ rate of attacks able to be misclassified by the classifier, here over $4,000$ attacks, none of the settings were able to exceed $2,500$ misclassified attacks.
Random modifications are not able to exceed $2,500$ misclassifications (regardless of the number of displacements, the step size or whether the training set is balanced or not) which corresponds to more than half the generated configurations but with a lower effectiveness than with our evasion attack.
The maximum number of misclassified configurations after random modifications starts from step size $t = 10,000$ regardless of the studied number of displacements.

Considering the validity of these configurations, results are similar to what can be observed in Figure~\ref{fig:attack_valid_acc}. The only difference is that the transition from $4,000$ to $0$ in the number of valid configurations is smoother and happens when $t$ is in $\left[0.01;100\right]$.
%Now, considering whether these configurations are valid or not, Figure~\ref{fig:rand_attack_valid_acc} shows that all random configurations are invalid \wrt to constraints from the VM starting at step size $1$ regardless of the displacements.
%Meaning that the potential of random displacements can never be exploited since the maximum of configurations that are misclassified starting from step size $\ge 10$ while all configurations are invalid starting step size equals to $1$.
%\begin{figure*}
%    \subfloat[Number of valid random attacks after 20 displacements \label{subfig:valid_rand_attack_20}]{%
%    \includegraphics[scale=.4]{./Images/result_valid_attack/random_attack/result_valid_random_20_disp_steps}
%    }
%    \subfloat[Number of valid random attacks after 50 displacements \label{subfig:valid_rand_attack_50}]{%
%    \includegraphics[scale=.4]{./Images/result_valid_attack/random_attack/result_valid_random_50_disp_steps}
%    }
%    \subfloat[Number of valid random attacks after 100 displacements \label{subfig:valid_rand_attack_100}]{%
%    \includegraphics[scale=.4]{./Images/result_valid_attack/random_attack/result_valid_random_100_disp_steps}
%    }
%    \caption{Number of valid random attacks on class -1}
%    \label{fig:rand_attack_valid_acc}
%\end{figure*}

\emph{Discussion:} Previous results lead us to state that the effectiveness of evasion attacks are superior to random modifications since i) evasion attacks are able to craft configurations that are always misclassified by the ML classifier while less than $2,500$ over $4,000$ generations will be misclassified using random modifications; ii) generated evasion attacks support a larger set of parameter values for which generated configurations are valid; iii) we were able to identify sweet spots for which evasion attacks were able to generate $4,000$ configurations that were both misclassified and valid.

\textbf{RQ1.4: Are attacks effective regardless of the targeted class?}
Previously, we generated evasion attacks from the class \textit{non-acceptable} and tried to make them acceptable for the ML classifier but is our attack symmetric? We thus configure our adversarial configurations generator so that it moves configurations from the class +1 (acceptable configurations) to the class -1 (non-acceptable).

\begin{figure*}
    \subfloat[Number of successful adversarial attacks after 20 displacements\label{subfig:adv_attack_20_nonacc}]{%
    \includegraphics[width=.44\linewidth]{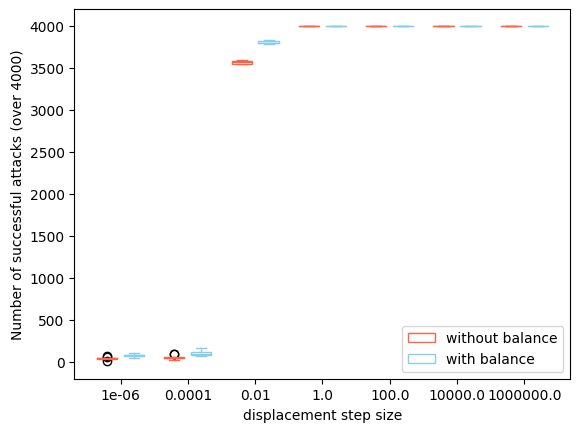}
    }
    \qquad
%    \subfloat[Number of successful adversarial attacks after 50 displacements\label{subfig:adv_attack_50_nonacc}]{%
%    \includegraphics[width=.45\linewidth,height=5cm]{./Images/result_exec_attack/adv_attack/result_50_disp_steps_nonacc_big}
%    }
    \subfloat[Number of successful adversarial attacks after 100 displacements\label{subfig:adv_attack_100_nonacc}]{%
    \includegraphics[width=.44\linewidth]{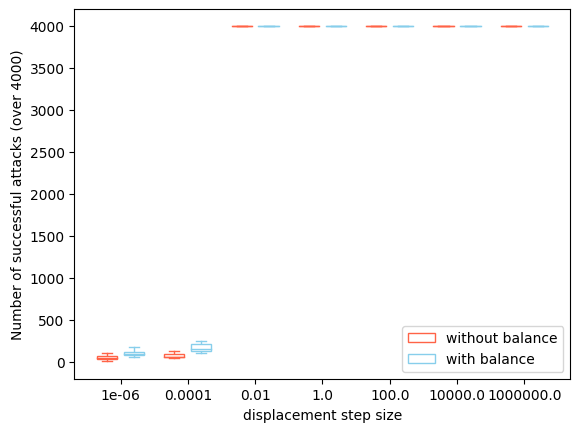}
    }
    \caption{Number of successful adversarial attacks on class \textit{non-acceptable}; X-axis represents different step size values $t$ while Y-axis is the number of misclassified adversarial configurations by the classifier; In orange and blue are respectively shown results when the training set is not balanced and when it is.}
    \label{fig:attack_success_nonacc}
\end{figure*}

Overall, the attack is symmetric: all generated adversarial configurations can be misclassified.
Figure~\ref{subfig:adv_attack_20_nonacc} shows that all generated configurations are misclassified when step size is set to $1$ or higher with a number of displacements of $20$ while, when the number of displacements is set to $100$ (see Figure~\ref{subfig:adv_attack_100_nonacc}), the step size can be set to $0.01$ or higher. These observations are the same regardless of the balance in the training set.

Regarding the adversarial configuration validity, a transition from $4,000$ to $0$ can still be observed.
However, when the number of displacements is set to $20$ or when the training set is balanced, the transition is abrupt and occurs when step size belongs to the $[100,10,000]$ range. With a higher number of displacements (\ie $50$ and $100$ and no balance), the transition is smoother but happens with smaller step sizes (\ie with $t$ in between $\left[ 0.01;100 \right]$. Adversarial configurations can be generated regardless of the targeted class even if targeting the least represented class seems promising.
%\vspace{-0.1cm}
%\begin{tcolorbox}
\SumupBox{
Our generated adversarial attacks are: $100\%$ effective (always misclassified, RQ1.1), do not depend on the target class (RQ1.4) and yield valid configurations (RQ1.2). In contrast, our random baseline was only able to achieve $62.5\%$ of effectiveness at best (RQ1.3). The balance in the training set does not affect these results and the targeted class affects show the same trends despite small differences (RQ1.4).}

\subsubsection{\textbf{RQ2: What is the impact of adding adversarial configurations to the training set regarding the performance of the classifier?}}

In our previous experiments, we only evaluated the impact of generated attacks in test set. Yet, some ML techniques (GANs) take advantage of adversarial instances by incorporating them in the training set to improve the classifier confidence and possibly performance. In our context, we want to assess the impact of our attacks when our classifier includes them in the dataset used during the training phase, especially with less ``aggressive'' (\eg small step sizes and a low number of displacements) configurations of the attacks.    

To do so, we allowed $20$ displacements in order to avoid configurations moving too far from their initial positions and we restrict the step size to every power of 10 in between $10^{-4}$ and $10^{1}$.
For each step size, we generate $25$ adversarial configurations that are added all at once in the training set, we retrain the classifier and evaluate it on the configurations that constitute the initial test set (without any adversarial configurations in it).
Every retraining process were repeated ten times in order to mitigate the effects of the random configuration selection and starting configurations. 
We also present results when the training set is balanced, in which case we have also augmented the test set to bring balance and to follow the same data distribution. In this case, the test set does not contain $4,000$ configurations but about $7,000$ in which $50\%$ of the configurations are considered acceptable and the remaining are considered non-acceptable. 

%Note that by adding configurations, results may be biased.
%A typical measure to evaluate a classifier is accuracy which is a ratio between the number of well classified configurations to the total number of configurations.
%By adding configurations, the denominator will increase naturally which might make artificially increase or decrease the value of the ratio.

\begin{figure}
    \centering
      \includegraphics[width=.9\linewidth]{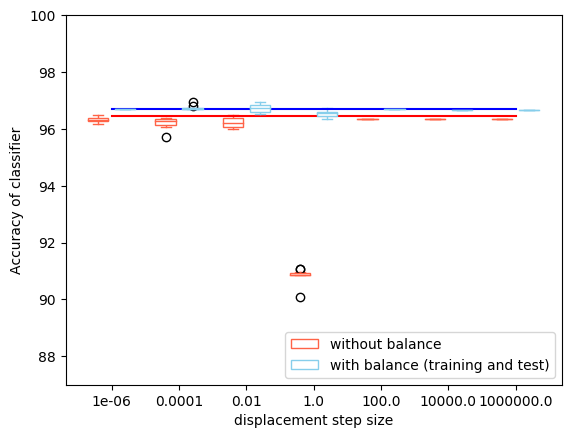}
    \caption{Accuracy of the classifier after retraining with $25$ adversarial configurations in the training set over a test set of $4,000$ configurations ($7,000$ configurations when the training set is balanced). In red are results when no balance are forced in the classes, in blue, both training set and test set are balanced. The initial accuracy of the classifier is represented by the horizontal line ($96.4562\%$ for the red line and $96.7143\%$ for the blue one). X-axis represents different step size values $t$ while Y-axis is the accuracy of the classifier (zoomed between $80\%$ and $100\%$).}
    \label{fig:retrain_25_adv_config}
\end{figure}

Figure~\ref{fig:retrain_25_adv_config} shows the accuracy of the retrained classifiers over a test set composed of $4,000$ configurations for the red part and $7,000$ configurations for the blue one.
%We first detail results when the sets are not balanced and then when they are.

The initial accuracy of the classifier was $96.4562\%$ over the same $4,000$ configurations and is shown as the horizontal red line. We make the following observations: i) using adversarial configurations in the training, even with low step sizes, tend to decrease the accuracy of the retrained classifier; ii) starting from step sizes of $1$, every run gives the same result.

Specifically, with step size equals to $10^{-4}$, the median of the boxplot is very close to the initial accuracy (\ie $96.4562\%$) of the classifier and the interquartile range is small suggesting that the impact of adding adversarial configurations into the training set is marginal.
%%with the interquartile range stretching a bit the accuracy around this value, sometimes making it better, sometimes worse.
Between $10^{-3}$ to $10^{2}$, the median is slightly decreasing and the interquartile range tends to increase.
At $t=10^{-1}$ the accuracy of the classifier drops to $91\%$, the adversarial configurations are specially efficient, forcing the ML classifier to change drastically its separation resulting in a lot of prediction errors. The last two step sizes shows that all the runs  with the same step size give the same results in terms of accuracy. Focusing on these runs, adversarial configurations had features with the same value: the amount of heat haze, of blur, of compression artefact or the amount of static noise of the $25$ adversarial configurations are all equal. All of these features are directly related to the global quality of images, and are key for the classifier accuracy. We explain the evolution of the classifier's accuracy as a combination of the contribution of the $8$ most important features and the constraints of the VM. For low step sizes ($t \in [10^{-4},10^{-2}]$), displacements are modest and therefore perturbations are very limited, though slightly observable. The sweet spot is at $t=10^{-1}$: the resulting displacement is important enough to move features values enough so that the associated configurations are moved effectively towards the separation and fool the classifier.  We computed the means and standard deviations between the initial and adversarial configurations and their difference witness the impact of adversarial configurations on the classifier. For larger values of $t$ (\ie $>$ to $10^{-1}$), these features lose their impact because their values are limited by constraints (so that they do not exceed the bounds).

In the case where training and test sets are balanced (in blue on Figure~\ref{fig:retrain_25_adv_config}), results follow the same tendency. Since most of added configurations to provide balance are well classified, we see that the accuracy is a bit higher than in the non-balanced.
Values remain close to the baseline, however, when $t = 1.0$, results are worse than for other executions as for the non-balanced setting.
Yet, we cannot conclude about the classifier robustness and more experiments should be conducted to take into account the fact the balanced and non-balanced datasets do not contain the same number of configurations.
%in order to discard the bias introduce by the way accuracy is computed with regards to the number of configurations in the test set.

%\begin{tcolorbox}
%Our attacks cannot improve the classifier's accuracy but can make it significantly worse: a single adversarial configuration over 500 can make the accuracy drop by 4\%. Successful attacks also exhibit the role of meaningful features intervening in the videos' acceptability.  VM constraints play a key role in the impact of the attacks and their interpretation. 
\SumupBox{Our attacks cannot improve the classifier's accuracy but can make it significantly worse: $25$ adversarial configuration over $500$ can make the accuracy drop by $5\%$.
Successful attacks also pinpoint visual features that do influence the videos' acceptability and that do make sense from the SPL perspective (computer vision).}

\section{Threats to validity}
\label{sec:threats}

%In the following, we describe potential threats that could mitigate the highlights we drew from our experiments.
\textbf{Internal threats.}
%\begin{itemize}
%    \item used only a limited set of parameters values to parameterize the attack
%    \item the choice of the attack technique
%    \item take into account constraints of the VM
%\end{itemize}
%\pt{Can you start "slowly" by explaining the threat? what is impacted by the threat? what's the possible consequence?}
%In the following, we discuss some choices that were made in our implementation, in the running of our experiments and that can mitigate the conclusions we drew from observations.
Choice of parameter values for your experiments may constitute a threat. The step size has been set to $10$ different power of $10$, we only used $3$ different number of allowed displacements (\ie $20$, $50$ and $100$).
From our perspective, using step sizes of $10^{-7}$ in a highly dimensional space seems ridiculously small while, on the contrary, using step sizes of $10^{4}$ are tremendously large which motivates our choice to not going over these boundaries.
However, the lower boundary could have been extended which might have affect results regarding \textit{RQ2}. Still, given the design of our attack generator, it is likely that performance of the classifier would never have increased.
Regarding the number of displacements, we could have used finer grained values. We sought a compromise between allowing a lot of small steps and a few big steps. Regarding the choice of evasion attacks, as presented in Section~\ref{sec:background}, several techniques exist. Evasion attacks showed interesting results and open new perspectives that we discuss in the Section~\ref{sec:discuss}.

We rely on centroids to deal with class imbalance (see Section~\ref{subsec:eval_proto}). The centroid method has pros and cons: centroids are easy and quick to compute, new configurations tend to follow the same distribution as they result in more densely populated clusters and on rare occasions, make clusters expand a little bit. However, new configurations may not be realistic, since they do not provide so much diversity -- centroids, by definition, lie in the middle of the cluster of points. Since our goal is only to limit imbalance in the available configurations, this technique is appropriate while maintaining the initial distribution of configurations. However, we are aware that other data augmentation techniques can be used. 

\textbf{External threats.}
%\begin{itemize}
%    \item only one case study $\rightarrow$ MOTIV but it is relevant and interesting (different nature of features)
%    \item only evaluate accuracy but it could work with others performance measures
%\end{itemize}
%The following discusses some aspects that might limit the use of our method to a more general context.
% \ma{we should better write this argument here}
 We only assessed our adversarial attack generator on one case study, namely MOTIV. Yet, MOTIV is a complex and industrial case exhibiting various challenges for SPL analysis, including heterogeneous features, a large variability space and non-trivial non-functional aspects. The x.264 encoder has been studied (\eg \cite{DBLP:conf/sigsoft/NairMSA17}) but is relatively small in terms of features (only 16 were selected), heterogeneity (only Boolean features) and number of configurations. This can nevertheless be a candidate for replicating our study. 
Our adversarial approach is not specific to the video domain and, in principle, applicable to any SPL. Generating adversarial configurations without taking into account all constraints of the variability model directly into the attack algorithm may threaten the applicability of our approach to other SPLs.
%Feature heterogeneity and the frequency of validity checks, make calls to SAT/SMT solvers unpractical. 
Calls to SAT/SMT solvers are unpractical due to feature heterogeneity and the frequency of validity checks.
Benchmarks of large and real-world feature models can be considered if we are only interested in sampling aspects \cite{DBLP:conf/se/KnuppelTMMS18, Siegmund:2017:AVM:3106237.3106251}. Finally, open-source configurable systems like JHipster \cite{DBLP:journals/ese/HalinNADPB19} can be of interest to study non-functional properties like binaries' sizes or testing predictions. We also considered accuracy as a the main performance measure. Accuracy is the standard measure used in the advML literature~\cite{barreno2006can,nelson2008,biggio2013poisoning,biggio2014security,biggio2014pattern,gan2014} to assess the impact of attacks.  

\section{Discussions}
\label{sec:discuss}

% of all the components of the SPL framework.
 % As we showed throughout this work, 
Adversarial configurations pinpoint areas of the configuration space where the ML classifier fails or has low confidence in its prediction. We qualitatively discuss what \emph{the existence of adversarial configurations suggests for an SPL} and to what extent the knowledge of adversarial configurations is actionable for MOTIV developers. 
% , we can wonder what does the existence of adversarial configurations suggest in regards to an SPL like the MOTIV generator. 
 %In reaction, developers of an SPL can have different attitudes. Taking a step backwards, we now qualitatively discuss what \emph{the existence of adversarial configurations suggests for an SPL} and to what extent the knowledge of adversarial configurations is actionable for MOTIV developers. 
% to exploit adversarial configurations. 

\textbf{\#1 Adversarial training.} A first reaction is simply to seek improvements of the ML classifier and making it more robust to attacks. 
% Several defenses have been proposed in the literature: manifold projections, stochasticity, prepossessing, etc.. 
% For investigating whether the idea works in the MOTIV context, we specifically investigate this question: \emph{Can we use AdvML techniques to improve the training set and thus the classifier?}
%\begin{itemize}
%    \item In traditional ML, adv config are added to the training set with the right label
%    \item the approach here is different and the goal of the study too
%    \item In MOTIV, the oracle is not well defined; asking humans to check is labour-intensive
%    \item One possibility is to control displacements to ensure that the config does not cross the separation
%    \item Future work: evalaute heuristics
%\end{itemize}
 Previous work on advML~\cite{biggio2018wild,barreno2006can, guo2017countering, dhillon2018stochastic, madry2017towards} proposed different defense strategies in presence of adversarial configurations. Adversarial training is a specific category of defense: the training sample is augmented with adversarial examples to make the ML models more robust against them. In our case study, it consists in applying our attack generator and re-inject adversarial configurations as part of the original training set.
% The most natural defense strategy seems to be the reactive defense which includes an adversarial configuration in the training set associated with the label given by the oracle\footnote{neither the one from the classifier nor the label from the original configuration as we did in this study}.
 We saw in \textit{RQ2} that, when adversarial configurations are introduced in the training set, even moderately agressive attacks affect the ML classifier performance. Our adversarial training is not adequate: our adversarial generator has simply not been designed for this defensive task and rather excels in finding misclassifications. It opens two perspectives. The first is to apply other, more effective defense mechanisms (manifold projections, stochasticity, prepossessing, \etc~\cite{biggio2018wild,barreno2006can, guo2017countering, dhillon2018stochastic, madry2017towards}). The second and most important perspective is to leverage adversarial ML knowledge for improving the SPL itself with ``friendly'' rather than malicious attacks, fooling the classifier is a mean to this objective. 
 
 % In addition, several aspects are different.
%MOTIV's oracle is an image quality assessment procedure~\cite{IQA} that is already an approximation of an other oracle (\ie human beings) making it fallible.
% 

\textbf{\#2 Improvement of the testing oracle.} The labelling of videos as acceptable or non-acceptable -- the testing oracle -- is approximated by the ML classifier.  If the oracle is not precise enough, it is likely that the approximation performs even worse. In the MOTIV case, oracles are an approximation of the human perception system which in turn could be seen as an approximation of the real separation between acceptable images and non-acceptable ones regarding a specific task. Object recognition should potentially work on an infinite number of input images which makes the construction of a ``traditional'' oracle (a function that is able to give the nature of every single input) challenging. Testing oracles for an SPL are programs that may fail on some specific configurations. Adversarial configurations can lead to ``cases'' (videos) for which the oracle has not been designed and tested for and may provide insights to improve such oracles.  %is loosely defined and too weak. After all, an ML classifier is an approximation of a testing oracle: in case the oracle is wrong so the ML classifier. The principle is that adversarial configurations can pinpoint "cases" (videos) for which the oracle has not been designed and tested for. 
% a real separation function (usually carried by the oracle) which is faster to compute and thus being easily scalable.

% In~\cite{temple:hal-01323446} we reported on the difficulty of engineering the oracle procedure.
%In summary, the fact that evasion attacks may work on a classifier (even highly trained and highly performant) can be the sign that the testing oracle itself is not sufficient.

% In short, the 

% An 
 % They also suggest efforts should be put in further exploring/sampling configurations in this area.
% If it is still not sufficient, maybe the problem is that the approximated oracle function is not strong enough.

MOTIV's developers may revise the visual assessment procedure to determine what a video of \textit{sufficient quality means}~\cite{IQA, temple:hal-01323446}. Adversarial configurations can help understanding the bugs (if any) of the procedure over specific videos (see Figure~\ref{img:examples_gen_vid}, page~\pageref{img:examples_gen_vid}). Based on this knowledge, a first option is to fix this procedure -- adversarial configurations would then act as good test cases for ensuring non-regression issues with the oracle. %Another possibility is to complement the original testing oracle with another oracle perhaps more strict and complete.
 In our context, one can envision to crowd-source the labelling effort with humans (\eg with Amazon Mechanical Turk~\cite{Mechanical_Turk}).   
 However, asking human beings to check whether a video is acceptable or not is costly and hardly scalable -- we have derived more than $4,000$ videos. Crowd-sourcing is also prone to errors made by humans due to fatigue or disagreements on the task. %Even if it was possible, due to the huge number of reviews to make, human beings could experience fatigue which would make them less accurate in the long term run.
To decrease the effort, adversarial configurations can be specifically reviewed as part of the labelling process. An open problem is to find a way to control adversarial displacements such that we are able to ensure that the generated adversarial configuration does not cross the ML separation. This level of control is left for future work.  Overall, the choice of the adequate testing oracle strategy in the MOTIV case is beyond the scope of this paper. Several factors are involved, including cost (\eg manually labelling videos has a significant cost) and reliability. %In this context, adversarial configurations are suited to assess and guide a possible improvement of the QA testing procedure. 

\textbf{\#3 Improvement of the variability model.} While generating adversarial configurations, SPL practitioners can gain insights on whether the feature model is under or over constrained. Looking at modified features of adversarial configurations (see \textit{RQ2}), practitioners can observe that the same patterns arise involving some features or combinations of features. Such behavior typically indicate that constraints are missing -- some configurations are allowed despite they should not be but it was never specifically defined as such in the variability model. Conversely, adversarial configurations can also help identifying which constraints can be relaxed.
% in the framework.

%\ma{Can we say a bit more here?}
%\pt{a toi de jouer!!}

%\textbf{About the diversity of adversarial configurations}
%\pt{merge "About the diversity of adversarial configurations" here (see comments in the LaTeX) if you think it's useful}

% With \textit{RQ2} we observed that features are quickly restricted with MOTIV and cannot be moved anymore.
% Obviously, it limits diversity of adversarial configurations but it may be a property of MOTIV.
% However, changing the order of modifications of the features and adapting the step size to the importance of features should help in avoiding the restrictions that provide features to move further.
% An other idea is to include in the training set and to retrain the classifier every time an adversarial configuration is created. Doing so, the retraining step will change the gradient function and thus the direction of modifications. Thus, it may foster the diversity in adversarial configurations.

\textbf{\#4 Improvement of the variability implementation.} Features of MOTIV are implemented in Lua \cite{Ierusalimschy:2006:PLS:1200583}. An incorrect implementation can be the cause of non-acceptable configurations either because of bugs in individual features or undesired feature interactions. In the case of MOTIV, we did not find variability-related bugs. We rather considered that the cause of non-acceptable videos was due to the variability model and that the solution was to add constraints preventing this.

\section{Related Work}
\label{sec:related_work}

%Our work aims to support quality assurance of SPLs through the use of ML techniques. 
% We propose to enhance their work by using adversarial ML techniques. 
% in order to retrieve more accurate constraints to be added to the Variability Model.
 Our contribution is at the crossroad of (adversarial) ML, constraint mining, variability modeling, and testing.

%In the context of \cite{temple:hal-01323446}, authors motivate their work using a Variability Model that could be used to derive test cases (video sequences) which in turn will test algorithms (in this case Computer Vision tracking algorithms).
%We propose to enhance their work by using adversarial ML techniques in order to retrieve more accurate constraints to be added to the Variability Model.
%Our contribution is thus at the crossroad of ML and adversarial ML, constraint discovery and variability management, software testing.

% \ma{Testing and SPL? We should discuss longer the "testing" as we are focusing on QA}

\textbf{Testing and learning SPLs.} Testing all configurations of an SPL is most of time challenging and sometimes impossible, due to the exponential number of configurations~\cite{thuem2014, MKRGA:ICSE16, Legay:2017:QRP:3023956.3023970,terBeek:2019:QVM:3302333.3302349, DBLP:conf/splc/VarshosazATRMS18, halin:hal-01829928}. ML techniques have been developed to reduce cost, time and energy of deriving and testing new configurations using inference mechanisms. For instance, regression models can be used to perform performance prediction of configurations that have not been generated yet~\cite{SGKA:ESECFSE15,guo2015,guo2013,DBLP:conf/isola/BeekFGS16,siegmund2013,DBLP:conf/sigsoft/OhBMS17} .
%They rely on a statistical analysis of values given to features of a configuration in order to retrieve correlations between the use of certain features or values and the performance of the resulting configuration.
% They rely on a statistical analysis of feature values in order to retrieve correlations between the performance of a configuration and activated features or their values.
% After correlations are computed, the statistical model can be used on new configurations estimate their performance without actually measuring them. 
% The model is precise if performance predictions are close to the real ones.
% 
 In~\cite{temple:hal-01323446, DBLP:journals/software/TempleAJB17}, we proposed to use supervised ML to discover and retrieve constraints that were not originally expressed before in a variability model. We used decision trees to create a boundary between the configurations that should be discarded and the ones that are allowed. In this paper, we build upon previous works and follow a new research direction with SVM-based adversarial learning.  
% The challenge is to restrict the usage of unwanted configurations without discarded too much configurations that could suit users' needs.
 % These needs are defined by a testing oracle (which can be human or automatic) which is used to label a first set of configurations.
 
%After that, the separation can be used on new configurations to predict their classes. 
% 
 Siegmund \etal ~\cite{Siegmund:2017:AVM:3106237.3106251} reviewed ML approaches on variability models. They propose THOR, a tool for synthesizing realistic attributed variability models. An important issue in this line of research is to assess the robustness of ML on variability models. Yet, our work specifically aims to improve ML classifiers of SPL. None of these bodies of work use adversarial ML neither the possible impact that adversarial configurations could have on the predictions.

\textbf{Adversarial ML} can be seen as set of security assesement and reinforcement techniques helping to better understand flaws and weaknesses of ML algorithms. Typical scenarios in which adversarial learning is used are: network traffic monitoring, spam filtering, malware detection~\cite{barreno2006can,biggio2013poisoning,biggio2014security,biggio2014pattern,biggio2013evasion,biggio2012poisoning} and more recently autonomous cars and object recognition~\cite{deeproad2018,deepXplore2017,canFoolBothGoodfellow2018,limitDLPapernot2016,accessorize2016,advInPhysical2016,physWorldAttacks2017}.
In such works, authors suppose that a system uses ML in order to perform a classification task (\eg differentiate emails as spams and non-spams) and some malicious people try to fool such classification system.
These attackers can have knowledge on the system such as the dataset used, the kind of ML technique that is used, the description of data, \etc The attack then consists in crafting a data point in the description space that the ML algorithm will misclassify. Recent works \cite{gan2014} used adversarial techniques to strengthen the classifier by specifically creating data that would induce such kind of misclassification.
In this paper, we propose to use a similar approach but adapted to SPL engineering: adversarial techniques may be used to strengthen the SPL (including variability model, implementation and testing oracle over products) while analyzing a small set of configurations. To our knowledge, no adversarial technique has been experimented in this context. % of SPL or variability-intensive systems.
\section{Conclusion}
\label{sec:conclusion}
%\gpe{to rewrite in the light of the previous changes...}
Machine learning techniques are increasingly used in software product line engineering as they are able to predict whether a configuration (and its associated program variant) meets quality requirements. ML techniques can make prediction errors in areas where the confidence in the classification is low. We adapted adversarial techniques on our MOTIV case and generated both successful and valid attacks that can fool a classifier with a low number of adversarial configurations and decrease its performance by $5\%$.  The analysis of the attacks exhibit the influence of important features and variability model constraints.  This is a first and promising step in the direction of using adversarial techniques as a novel framework for quality assurance of software product lines. As future work, we plan to compare adversarial learning with traditional learning or sampling techniques (\eg random, t-wise). Generally we want to use adversarial ML to support quality assurance of SPLs. 
\balance
\bibliographystyle{ACM-Reference-Format}
\bibliography{references,macher}

\end{document}